\documentclass[10pt,preprint2]{aastex}

\newcommand{\etal}{et~al.}

\newcommand{\water}{$\mbox{H}_{2}\mbox{O}$ }

\newcommand{\degrees}{$^\circ$}
\newcommand{\kms}{kms$^{-1}$}

\shorttitle{Scintillation in Circinus}
\shortauthors{McCallum \etal}

\begin{document}

\title{Scintillation in the Circinus Galaxy \water megamasers}


\author{Jamie N. McCallum\altaffilmark{1}, Simon P. Ellingsen}
\affil{University of Tasmania, School of Maths and Physics, Private Bag 21, Hobart, TAS 7001, Australia}\email{jmccallu@utas.edu.au, sellings@kerr.phys.utas.edu.au}
\author{David L. Jauncey, James E. J. Lovell}
\affil{ATNF, PO Box 76, Epping NSW 1710, Australia}
\email{David.Jauncey@csiro.au, Jim.Lovell@csiro.au}
\author{Lincoln J. Greenhill}
\affil{Harvard-Smithsonian Centre for Astrophysics, 60 Garden St, Cambridge, MA 92138, USA}
\email{lincoln@cfa.harvard.edu}

\altaffiltext{1}{Affiliated with the Australia Telescope National Facility}


\begin{abstract}
We present observations of the 22 GHz water vapor megamasers in the Circinus galaxy  made with the Tidbinbilla 70~m telescope. These observations confirm the rapid variability seen earlier by \citet{Gr97}. We show that this rapid variability can be explained by interstellar scintillation, based on what is now known of the interstellar scintillation seen in a significant number of flat spectrum AGN. The observed variability cannot be fully described by a simple model of either weak or diffractive scintillation. 

\end{abstract}





\section{Introduction}
Water vapor megamaser emission was first detected from the Circinus galaxy by \citet{GW82}, making it the second closest Seyfert galaxy, after NGC 4945, to host water vapor megamaser emission. Recent VLBI observations show the emission arises in a thin warped disc with a clumpy, wide-angle outflow subtending an angle of $\sim$ 80 mas which corresponds to a linear region of $\sim$ 2 pc at a distance of 4.2 Mpc. The nucleus of the galaxy hosts a 1.7 $\pm$ 0.3 $\times 10^6$ M$_{\odot}$ black hole  which is also a powerful X-ray source \citep{Gr03}.

\citet{Gr97} observed very rapid variations in several of the strong maser lines, with timescales as short as a few minutes in one line. They discussed both intrinsic variability and diffractive interstellar scintillation (DISS) as potential mechanisms, and concluded that both were possible although favouring DISS. Both mechanisms placed limits on the physical conditions of the masers, with intrinsic variations requiring short gain paths of 1-10 AU, and DISS requiring angular sizes $\leq$ 1.75 $\mu$as ($\leq$ 7 AU at a distance of 4.2 Mpc). The implied brightness temperatures are in excess of $10^{16}$ K. Similar conclusions were reached by \citet{Maloney02}.

Over the last five years unequivocal evidence has accumulated that weak interstellar scintillation (WISS) is the principal mechanism responsible for the intra-day variability (IDV) observed at cm-wavelengths in many compact, flat-spectrum AGN (e.g Jauncey \etal  (2003) and references therein). In the light of these recent developments we show here that the very rapid variability seen in Circinus is also well described by ISS.

\section{Observations}


We present data from three epochs, one observation made with the Parkes 64~m telescope on DOY 289 of 1995 and two observations made in 1996 with the Tidbinbilla 70~m. The Parkes observation has been previously analysed and published by Greenhill \etal (1997). The details of the observation and data reduction are not reproduced here.

The Tidbinbilla 70~m observations were made at high elevations in left circular polarization on 1996 March 4 and March 28 (DOYs 124 and 148) using a 20 MHz bandpass with 1024 spectral channels giving a channel spacing of 19.5 KHz or 0.26 \kms. The bandpass was centred on 537 \kms with a velocity range of 402-672 \kms on DOY 124, and centred on 355 \kms with a velocity range of 220-490 \kms on DOY 148 (Heliocentric frame of reference). Circinus was observed in position switching mode with groups of eight 45-second integrations on- and off-source. The total duration of the observations was 2.8 hours on DOY 124 and 1.9 hours on DOY 148. 

\begin{figure}
        \epsscale{.80}
        \plotone{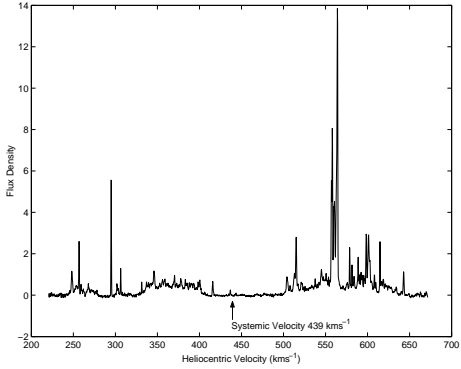}
        \caption{Average spectrum of the Circinus megamasers, made using the Tidbinbilla telescope. The two observations made on DOYs 124 and 148, of the red- and blue-shifted emission respectively, have been amalgamated. The velocity range covered by both observations is between 402 and 490 \kms.}\label{fig:TidSpec}
\end{figure}

The data on both days were calibrated assuming a constant 150 Jy system temperature so there may be small, slow trends due to atmospheric effects and slow changes in gain with elevation. The weather on both occasions was fine and the measured gain-elevation curve shows that such effects will be both slow and small. Also, there is no significant correlation between the observed variations of different maser features which indicates that our calibration is adequate. The measured RMS noise level in a spectral channel in a 45 second integration was 0.11 Jy and 0.17 Jy on DOY 124 and 148 respectively. The average spectrum is shown in Figure \ref{fig:TidSpec}.

Characteristic time-scales for variability, T$_{\rm{char}}$, were calculated from the normalized auto-correlation function (ACF) of the light curves for both the Parkes and Tidbinbilla data, using the discrete autocorrelation function \citep{EK88}. Following Rickett \etal (2002), we measure the lag at which the ACF falls to 0.5 as the estimate of T$_{\rm{char}}$. The velocities of the Parkes data have been re-calculated as it appears that these were previously in reference to the LSR despite being labeled as heliocentric in \citet{Gr97}. 

There are some difficulties in calculating the ACF of the data, due to the regular observing gaps. This is because the number of points for a lag interval decreases as the lag time approaches the gap time. The regular sampling pattern, with equal integration times on- and off-source and losses due to calibration and slewing overheads means that there are no data for some lags. Using the discrete correlation function, we have excluded those lags that were computed using less than a third of the available data.




\section{Results and Discussion}

Table \ref{tab:Var} summarises the measured properties of the well-observed lines in the data. The velocity and FWHM of the lines are taken from the fitted Gaussian profiles as is the mean flux density ($<\rm{S}>$) and standard deviation of the variariability ($\sigma$). The modulation index, $\mu$, was calculated by $\sigma / <\rm{S}>$. The characteristic timescale, as defined above, is included together with the estimated depth of the ACF at its first minimum. The listed errors were calculated using the estimation error formula of Appendix A of \citet{RKCJ02}. The extremely large errors are due to the short observing time and illustrate the need for longer observations.

\begin{figure}
\epsscale{.80}
\plotone{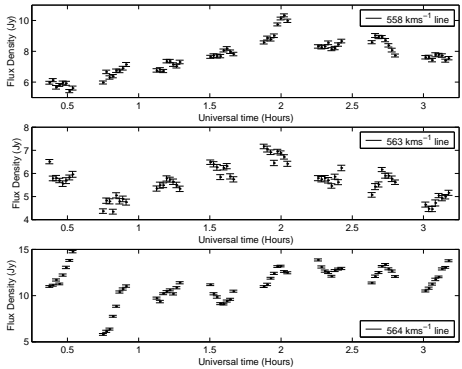}
\caption{Light curves of the three strongest lines, day 124 of 1996 from Tidbinbilla. A range of timescales is apparent in the data, with short-timescale, low amplitude variations overlaid on a much larger, slower trend in the 558 \kms line. The strongest line, at 564 \kms, is observed to more than halve its flux density in less than 10 minutes. This is the largest proportional change observed in this set of observations.}\label{fig:RedVar}
\end{figure}

\begin{figure}
\epsscale{.80}
\plotone{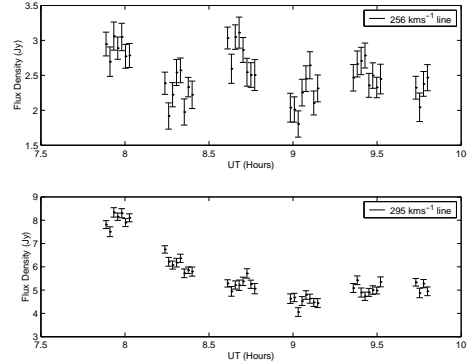}
\caption{Light curves of the two strongest lines observed on day 148 of 1996 from Tidbinbilla. The upper panel shows the light curve of the 256.79 \kms line and the lower panel the light curve of the 295 \kms line. The two lines display very different timescales of variability. While the 256 \kms line completes two quasi-periods, the 295 \kms line shows perhaps half of one.}\label{fig:BlueVar}
\end{figure}

The light curves presented in Figures \ref{fig:RedVar} and \ref{fig:BlueVar}, as well as the original Parkes light curves \citep{Gr97}, clearly show the rapid variability in the stronger lines. It is apparent that there is no significant cross-correlation in the Tidbinbilla data between the variations at the different line frequencies, showing that the variations cannot be due to systematic effects, miss-pointing or atmospheric effects for example. Moreover, different lines show quite different time scales, with the 295 \kms line on DOY 148 showing much slower variations than the 257 \kms line. There is, however, some evidence for rapid, low amplitude variations superimposed on the slow variations in the 295 \kms line. 

There are two strong lines present in both the Parkes and Tidbinbilla observations (the 295 and 557~\kms lines). In both cases, they are much weaker in the Tidbinbilla observations, having roughly halved their flux densities. The modulation indices, timescales and FWHM (Full-width half-maximum of the fitted Gaussian profile) of the masers have also changed. In the DOY 124 observations of the 557 \kms line, it has narrowed considerably and has a slightly faster timescale while the modulation index is very similar. For the 295 \kms line, its FWHM is unchanged while the modulation index has doubled and the timescale has significantly increased. A prediction of ISS theory is that the characteristic timescale of variability should change throughout the year as the motion of the Earth around the Sun changes the relative speed of the ISM, across the line of sight \citep{MJ02}. This should then cause the variability of both lines to speed up or slow down. The observed timescale changes, where one line slows down dramatically and the other speeds up tells us that source structure is affecting the scintillation and may be evolving on timescales of a few months. The observed line narrowing in the 557 \kms line is indicative of this and previous observations have indicated that the average lifetime of an individual maser feature is $\sim$ 1 year.

\subsection{Weak scintillation}
The quasi-sinusoidal variations of the light curves in Figures \ref{fig:RedVar} and \ref{fig:BlueVar} plus those in Greenhill et al's Figure 2, are remarkably similar in amplitude and time scale to the variations seen at 4.8 and 8.4 GHz in the extragalactic radio sources PKS~0405--385 \citep{KCJWWNRT97}, PKS~1257--326 \citep{BJLTKCMTRC03} and J1819+3845 \citep{DTdB2000}. Weak interstellar scintillation has been demonstrated as the principal cause of these rapid variations in each of these sources, PKS~0405--385 \citep{RKCJ02}, PKS~1257--326 \citep{BJLTKCMTRC03} and J1819+3845 \citep{DTdB03}, so it follows that WISS should also be considered as potentially responsible for the variability of the lines in Circinus. 

These three ``fast" AGN scintillators exhibit such rapid variability because they undergo weak scattering and have been found to be behind nearby scattering screens. The evidence for weak scattering is well established; the flux density variations are very broad band and are highly correlated at 5 and 8 GHz, and each of these sources shows much slower variations at 1.4 and 2.4 GHz in the strong scattering regime.

\begin{figure}
\epsscale{.80}
\plotone{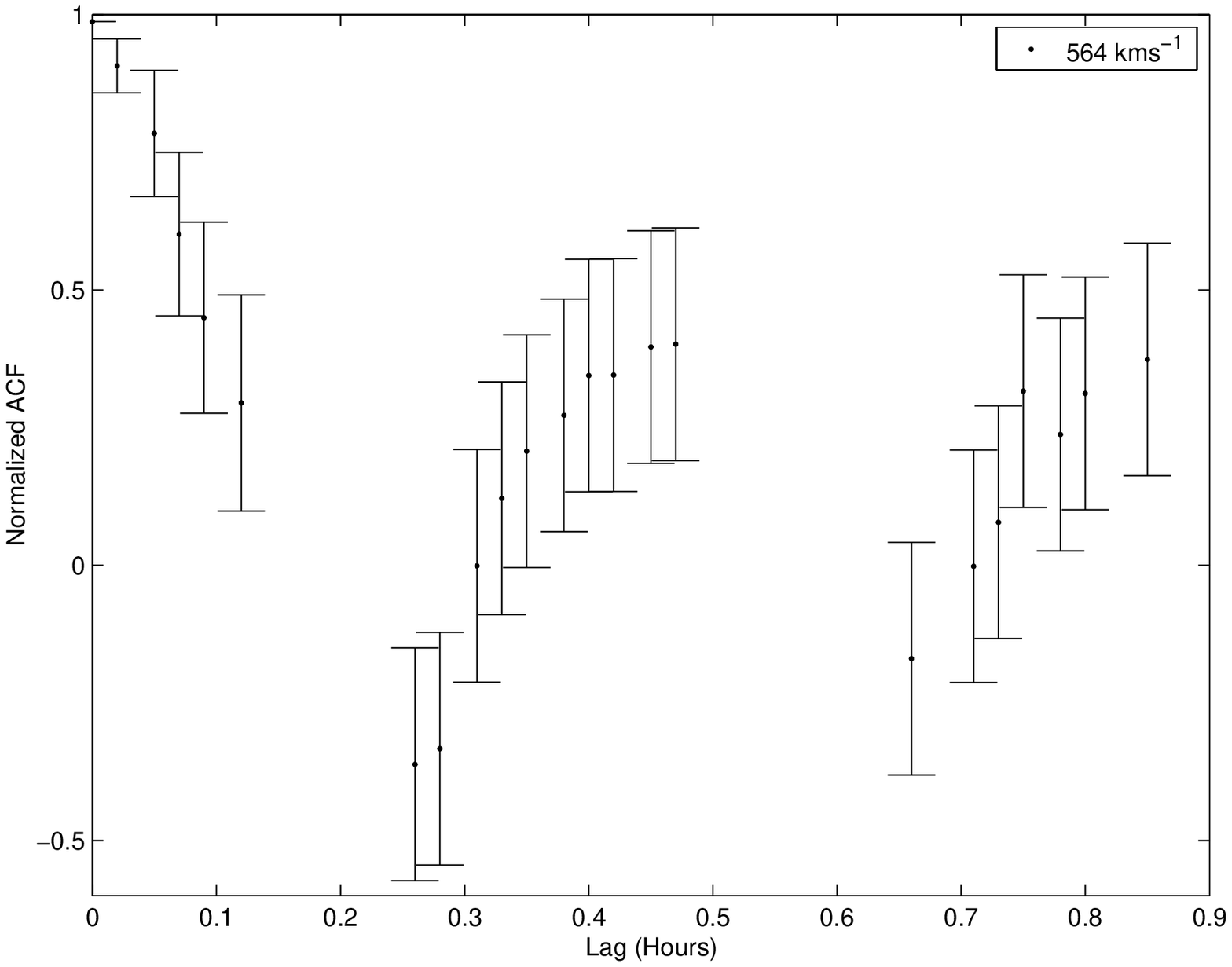}
\caption{Autocorrelation function of the 564 \kms line. This was computed using the discrete autocorrelation function as described in \citet{EK88}. The lags are binned in 40 second intervals. Data are indicated by dots and verical bars indicate the one $\sigma$ estimation error, following \citet{RKCJ02}. The large estimation errors are due to the short observing interval. Missing lags correspond to where the sampling function has left too few data to make a reliable estimation of the ACF. The ACF shows a deep first minimum and a strong quasi-periodic character.}\label{fig:DCF}
\end{figure}

Another strong similarity between the well established AGN weak scintillation and the rapid variations in the Circinus masers is the presence of a deep first minimum in the ACF. This is clearly illustrated in Figure \ref{fig:DCF}, which is the ACF of the strongest line in the DOY 124 observations. In the case of PKS~0405--385, the deep first minimum has been shown to indicate scattering in an anisotropic ISM \citep{RKCJ02}.

The evidence for nearby screens for the fast scintillators is also compelling. The linear size of the first Fresnel zone is proportional to (D/$\nu$)$^{0.5}$, and hence the corresponding time-scale for traversing this zone is proportional to (D/$\nu$)$^{0.5}$ $\times \rm{V}_{\rm{ISM}}^{-1}$, where D is the screen distance, $\rm{V}_{\rm{ISM}}$ is the speed of the interstellar medium (ISM) and $\nu$ the frequency. Thus for the measured values of $\rm{V}_{\rm{ISM}}$  for these sources, $\sim$ 30 km/s, and for the observed frequencies of $\sim$ 5 - 8 GHz where weak scattering predominates in AGN, screen distances of $\sim$ 15 to 30 pc are implied for PKS~0405--385 \citep{RKCJ02}, PKS~1257--326 \citep{BJLTKCMTRC03} and J1819+3845 \citep{DTdB02}.

For Circinus, and following Walker (1998);

T$_{\rm{char}}$ = 1.2 $\times$ 10$^6 \times$ (V$_{\rm{ISM}})^{-1} \times$ (D/$\nu$)$^{0.5}$,

where T$_{\rm{char}}$ is the characteristic time-scale of the observed quasi-sinusoidal variations (in seconds). Substituting T$_{\rm{char}}$ = 700 seconds, typical of the time-scales in Table 1, V$_{\rm{ISM}}$ = 50 \kms (Rickett et al., 1995), yields a screen distance, D, of 20 pc. If the screen is moving at the $\sim$ 30 \kms of the local standard of rest (LSR) and the values determined for the fast scintillators, then the screen distance is reduced to 7 pc. These screen distance estimates are remarkably similar to those found for the three fast AGN scintillators. At 22~GHz, the angular size of the first Fresnel zone is 12 $\mu$as for a screen distance of 20 pc. This is much larger than the expected size of the individual maser features. Galactic masers observed using VLBI subtend milli-arcsecond sizes at kpc distances which implies micro-arcsecond sizes at Mpc distances as the physical sizes are likely to be comparable. The Tidbinbilla observations show that all of the lines that have a sufficient SNR appear to exhibit rapid variations, as might be expected from the large angular size of the first Fresnel zone. This, in turn, allows an estimate of the transverse extent of the scattering screen. The overall angular scale of the maser emission from the galaxy is $\sim$ 80 mas (Greenhill et al., 2003), which, at 20 pc projects to a linear size of 2.4 $\times 10^8$ km. If all of the disc and outflow lines scintillate rapidly, then this represents a lower limit to the size of the nearby scattering screen. Circinus is the only source which has such close multiple scintillating components, so this is the first time that such a size estimate can be derived.

For this to be weak scattering requires a transition frequency between and weak scattering ($\nu_0$) of somewhat less than 22~GHz. The mean modulation index is 0.19, so that following Walker's (1998) equation 6, and assuming quite reasonably that the maser features are unresolved on a 12 $\mu$as scale, yields a transition frequency of 7~GHz, little higher than the 3 to 5 GHz found for the AGN, albeit at higher Galactic latitudes. We are really only looking here at the scattering that is taking place at the nearby screen. Circinus lies at a low Galactic latitude of -3.8\degrees and so the radiation is passing through a dense and turbulent region of the Galaxy. Models of the Galactic electron distribution such as Cordes and Lazio's NE2001 model\footnote{Available at \it{http://rsd-www.nrl.navy.mil/7213/lazio/ne\_model/}} \citep{CL03} predicts a transition frequency of 56~GHz for the line of sight to Circinus, placing the 22~GHz water maser radiation in the strong scattering regime. Also, as this takes the entire path length through the Galactic plane into account, the screens are likely to be at a distance of hundreds or thousands of parsecs. We will address the possibility of diffractive scintillation in the next section, but these distant screens are likely to also cause refractive scintillation. This will have a characteristic timescale of several hours and a relatively large modulation index, given by $\mu = \xi^{-\frac{1}{3}}$, where $\xi$ is the scattering strength \citep{W98}. For the predicted transition frequency of 56~GHz, this results in a modulation index of $\sim$ 0.6. 

There is a clear observational bias present in our data towards detecting rapid variations on a time scale of minutes to hours but missing those variations with time-scales of many hours. Both our data and the well-sampled part of Greenhill \etal's data span no more than 3 to 4 hours. However, both the 1995 Parkes and the 1996 Tidbinbilla data show evidence of longer-term variations with time-scales of at least several hours. It would appear highly likely, if WISS is the mechanism responsible for these variations, longer duration observations will uncover longer time-scales, of hours to days, which would be caused by scattering at a more distant screen(s). This would be the first time that scattering from multiple screens has been observed.

Using a local screen, we can explain a many of the observed characteristics of the variability in the Circinus megamasers. However, one problem is that we do not see all the lines varying on a single timescale which is expected for weak scattering. The large size of the first Fresnel zone means that source structure is unlikely to be responsible for the range of observed timescales unless both the source and scintillation pattern are highly anisotropic. Axial ratios of 6 or more have been suggested for scintillating sources such as PKS~0405--385 and J1819+3845 and VLBI images of Galactic masers often show elongated structures \citep{IDS02}. Another possibility is that refractive scattering at some distant screen has broadened the image of the maser sources to a size comparable to the first Fresnel zone of the local scattering screen. Anistropic structures will create distinctive patterns in the annual cycle.

It would appear that the rapid variability observed in the maser lines in the Circinus galaxy is indeed consistent with weak interstellar scintillation, based on the striking similarities with the variations observed in the other three fast AGN scintillators. Like the fast AGN scintillators, rapid variations imply the presence of a nearby scattering screen, and the presence of such a nearby screen is now more readily acceptable given those already found. These screens for the three AGN and Circinus are spread widely across the sky and suggest the possible presence not so much of a ''local bubble" (Bhat, Gupta \& Rao, 1998) as the remains of a burst bubble.

\subsection{Diffractive scintillation}
Because of the rapidity of the intensity variations in Circinus, \citet{Gr97} and Maloney (2002) both suggested diffractive interstellar scintillation (DISS) in the turbulent ISM, rather than the weak scattering in a very nearby screen that we have considered above. With DISS it is feasible to produce such short time scales for the maser lines at 22~GHz, using a distant screen. The diffractive timescale is given by t$_{\rm{diff}} = \rm{t}_{\rm{F}} / \xi$ and for a scattering strength of 5 (as predicted by the NE2001 model \citep{CL03}) and a screen at 1 kpc, the corresponding diffractive timescale is $\sim$ 1000 s. 

However, there are several problems in applying diffractive scintillation to the Circinus megamasers. Firstly, as noted by Greenhill et al. and Maloney, the observed $\sim$ 700 second time-scale observed at 22~GHz implied a time scale that was significantly shorter than found in pulsars at comparable Galactic latitudes at lower frequencies. The diffractive timescale varies with frequency as $\frac{\nu}{\nu_0}^{1.2}$ which means that the diffractive timescale at 1~GHz is a factor of 40 less than at 22~GHz \citep{W98}. The predicted timescale is therefore on the order of 20 seconds, which is comparable to some of the most rapidly scintillating pulsars \citep{C86}.

However, Circinus is extragalactic while most pulsars are within the Galaxy and as such experience less scattering. Also, the combination of steep pulsar spectra, narrow diffractive bandwidths and intrinsic pulse variations makes it extremely difficult or impossible to observe very short timescale intensity variability in most pulsars. Intrinsic pulse variations have very large modulation indices (Kramer \etal 2003 reports a modulation index of 1.21 for intrinsic variations at 1.4~GHz, in the pulsar B1133+16) and the variations in the pulses are uncorrelated. Looking for ISS involves averaging over a large number of pulses and then examining the variability of the averaged data. The averaging time has to be in excess of a hundred pulses to remove the effect of this intrinsic variability and hence timescales of the order of 10 seconds or less cannot be easily measured directly. Estimating the strength of the scattering could be attempted by measuring dynamic spectra of the two high dispersion measure (DM) pulsars (see below) close to the line of sight to the Circinus galaxy.

In modeling the variability of the Circinus megamasers as due to diffractive scintillation, we note that the masers have a modulation index of $\sim$ 0.20, rather than unity as expected for diffractive scintillation. This implies that the source has structure on scales greater than the diffractive limit. This extended structure acts as a spatial filter, reducing both the speed and amplitude of the variations. 

\begin{figure}
\epsscale{1}
\plotone{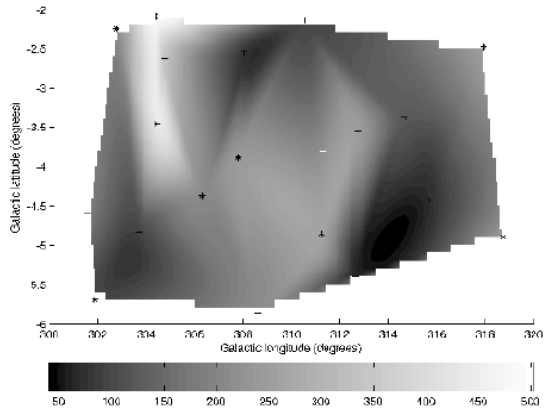}
\caption{This is a smoothed map of DM for pulsars along the Galactic plane near Circinus. The position of the Circinus Galaxy is indicated with a white star while the individual pulsars are indicated with black ones. The data was smoothed using two-dimensional spline interpolation. The data has been taken from the ATNF pulsar database}\label{fig:DM}
\end{figure}

In quenched diffractive scintillation, the modulation index is reduced to $\mu = \theta_{\rm{S}} / \theta_{\rm{diff}}$ and the timescale is increased by a factor of $\theta_{\rm{S}} / \theta_{\rm{diff}}$ so that the dominant lengthscale is now the linear size of the source, as projected onto the scattering screen. Using this model the observed timescale of a maser feature would be proportional to its angular size. Taking the most rapidly varying feature with its 700 seconds timescale and modulation index of 0.20, we can apply plausible estimates of source size and ISM velocity to investigate the screen. Using an ISM speed of 50 \kms and a source size of 1 $\mu$as, we obtain a screen distance of $\sim$ 230 pc, with a scattering strength of 18 and a transition frequency of 120~GHz. Decreasing the assumed angular size of the masers increases both the implied screen distance and scattering strength. For a 0.25 $\mu$as maser (which corresponds to a linear size of $\sim$ 1 AU at Circinus), the implied screen distance is $\sim$ 1 kpc with a scattering strength of $\sim$ 40 and a transition frequency of $\sim$ 200~GHz. While this is extremely high and well above that predicted by the NE2001 model for the line of sight to Circinus, it is characteristic of lines of sight at lower Galactic latitudes. We have inspected the ATNF pulsar database\footnote{\it{http://www.atnf.csiro.au/research/pulsar/psrcat/}} and it is worth noting that the two pulsars closest to Circinus both have quite high DMs $\sim$ 250 pc/cm$^3$, which is higher than that of other pulsars at comparable Galactic latitudes. We have used the ATNF pulsar database to produce a DM map of the Galactic plane near to the line of sight to Circinus (Figure \ref{fig:DM}), which shows that the line of sight to Circinus is associated with high DMs. This suggests that either the pulsars in this region are more distant than some of the other, less scattered pulsars or that there is some enhanced scattering occuring along that line of sight, or perhaps both. At present, none of the pulsars used in the DM map have had their distances measured independently. H$\alpha$ images of the same region show strong emission nearby, which is suggestive that enhanced scattering may be present.

Another problem with applying a diffractive scintillation model is that the expected anticorrelation between timescale and modulation index is not apparent in the data. Figure \ref{fig:BlueVar} illustrates this well with the very large amplitude flux variation in the 295 \kms line occurring on a very long timescale, contrasted with the rapid, lower amplitude variations in the 256 \kms line. This may be an effect of source and/or screen anisotropy which is suggested by the deep first minima of the ACFs, which will produce a range of  timescales depending on the relative orientations of the  axes of anisotropy and screen velocity.

\section{Conclusions}
We have shown that the variability of the Circinus megamasers has continued since its first detection and that it can be explained through ISS. The Circinus megamasers offer a unique opportunity to study both the interstellar scattering medium and the structure of extragalactic megamasers at high angular resolution. 

There are several experiments that can be made to confirm that scintillation is the cause of the variability and identify whether is due to a weakly scattering local screen or a strong distant one. 

Monitoring the variability of the Circinus megamasers will allow us to test for the presence of an annual cycle in its characteristic timescale. The annual cycle is created by the motion of the Earth around the Sun and can be used to measure the peculiar velocity of the scattering material, estimate the extent and orientation of anisotropies in the medium and to create a model of the source itself as has been accomplished for PKS~0405--385 \citep{RKCJ02}, J1819+3845 \citep{DTdB03} and PKS~1257--326 \citep{Bignall04}. Both of our ISS scenarios expect that the scattering medium will be anisotropic and that source structure is directly involved in the timescale of variability. This means that different maser features may display different annual cycles as a result of their structure. In the case of a local screen, its intrinsic velocity is likely to be comparable to local standard of rest, producing a clear annual cycle with speeds $\leq$ 50 \kms. A more distant screen may have some significant intrinsic velocity which will make the annual cycle harder to detect as the difference between the high and low speeds becomes proportionally less. The relative transverse speed of the ISM can also be directly measured by observing the time delay between the scintillation patterns recorded at two widely separated telescopes as has been done for PKS~0405--385 \citep{Jauncey0405}, PKS~1257--326 \citep{Bignall04} and J1819+3845 \citep{DTdB02}, using intercontinental baselines. A direct measurement of the ISM velocity is an extremely valuable constraint in producing models of anisotropic structure. 

These experiments will not conclusively distinguish between a local or distant screen model. One test would be to investigate longer timescale variability, which is predicted by both models as a result of refractive scintillation at some distant screen. In the local screen scenario, the refractive timescale would be decoupled from that of the rapid variations and while in the distant screen, both the diffractive and refractive variations would display the same annual cycle. The masers should all be point sources to refractive scintillations and so the timescale and modulation index is expected to be the same for all the masers. Also, the pulsars near to Circinus provide another probe of the ISM. If characteristic timescales and diffractive bandwidths can be measured from dynamic spectra then these can be used to make estimates of the screen strength and distance. 

We have shown that the presence of variability in the Circinus megamasers can be explained as due to ISS and have presented two models. At this stage, both require unusual, although not unprecedented, conditions to explain some characteristics of the Circinus megamasers. The local screen model using weak scintillation can explain the observed variability using a local, weakly scattering screen with a distance $\leq$ 20 pc and $\nu_0 \sim 7$. This is comparable to what has been observed in the most rapidly scintillating AGN sources. The appearance of the light curves is also strongly reminiscent of those seen in scintillating AGN. However, it does have some problems in explaining the observed range of timescales, as the angular size of the first Fresnel zone vastly exceeds that of any realistic source, unless there is also scattering by a more distant screen. Also, the low Galactic latitude of Circinus and the predictions of models of the Galactic electron distribution suggest that strong scattering and distant screens should be involved. The distant, strongly scattering screen model we have proposed satisfies most of our observations and fits with the predictions of electron density models. However, it predicts an anti-correlation between modulation index and timescale which we do not observe. Further observations will allow us to refine our models and to resolve some of our present uncertainties.

\acknowledgments
\begin{centering}
\bf{
Acknowledgements}
\end{centering}

The ATNF is funded by the Commonwealth Government for operation as a National Facility by CSIRO. Financial support for this work was provided by the Australian Research Council. We also wish to thank Edward King for making the Tidbinbilla observations and developing the software to read in the data, and to our referee for their extremely helpful comments.

\clearpage

\clearpage

\begin{deluxetable}{cccccccccccc}
\tabletypesize{\scriptsize}
\tablecolumns{9}
\tablecaption{Variability parameters of the observable lines\label{tab:Var}}
\tablehead{\colhead{Velocity} & \colhead{FWHM} & \colhead{$<\rm{S}>$} &  \colhead{$\sigma$} & \colhead{$\mu$} & \colhead{$\rm{T}_{\rm{char}}$ } & \colhead{First min.} \\
\colhead{(\kms)} & \colhead{(\kms)} & \colhead{(Jy)} & \colhead{(Jy)} & \colhead{} & \colhead{(Hours)} & \colhead{Depth} \\}
\startdata
\cutinhead{Parkes, day 289 of 1995}
295.03 $\pm$ 0.02 & 0.59 $\pm$ 0.04 & 11.97 &  1.22 & 0.1  & 0.07 $\pm$ 0.01  & -0.4 $\pm$ 0.2\\
 556.97  $\pm$  0.05  &  1.32   $\pm$ 0.16  & 11.36  &  1.51  &  0.13 &       0.4 $\pm$ 0.2 & No min.  \\
  562.05   $\pm$ 0.03  &  1.04   $\pm$ 0.08 &  18.03 &   3.64 &   0.20 &   0.3 $\pm$ 0.2    &  -0.2 $\pm$ 0.3 \\
\cutinhead{Tidbinbilla, day 124 of 1996}
   514.74 $\pm$   0.04 &   1.28 $\pm$   0.08 &   2.41 &   0.34 & 0.42   & 0.4 $\pm$ 0.2 & -0.4 $\pm$ 0.4 \\
  557.03 $\pm$   0.03   & 0.86  $\pm$  0.06   & 4.93 &   0.75  &  0.15    &  0.3 $\pm$ 0.1  &     -0.5 $\pm$ 0.4 \\
  558.13 $\pm$ 0.02 & 0.87 $\pm$ 0.04 & 7.66 &  1.14 & 0.15&  0.5 $\pm$ 0.3 & No min. \\
  560.3 $\pm$ 0.3\tablenotemark{a}& 1.9 $\pm$ 0.4 & 4.1 & 0.4 & ... & ...& ...\\
 563.24 $\pm$   0.05  &  1.88  $\pm$  0.07 &   5.67 &  0.69   & 0.12 & 0.26 $\pm$ 0.13  &     -0.8 $\pm$ 0.3\\
  564.22  $\pm$  0.05   & 0.84  $\pm$  0.05  & 11.32 &  1.84  &  0.16 &   0.09 $\pm$ 0.02  &   -0.4 $\pm$ 0.2\\
  578.46  $\pm$  0.03 &   0.78  $\pm$  0.11 &   1.59  & 0.37  &  0.23    & 0.16 $\pm$ 0.06 &   -0.8 $\pm$ 0.3 \\
  598.17  $\pm$  0.03   & 1.18  $\pm$  0.13   & 2.26 &   0.23  &  0.10 &     0.16  $\pm$ 0.06        &   -0.2 $\pm$ 0.3\\
  614.53 $\pm$  0.03   & 0.82   $\pm$ 0.08 &   1.96 &  0.24  &  0.12 &  0.09 $\pm$ 0.02  &   -0.8 $\pm$ 0.2\\
  642.65  $\pm$  0.09  &  0.96  $\pm$  0.18   & 0.91 &    0.25  &  0.27 &    0.24 $\pm$ 0.1  &  -0.2 $\pm$ 0.3 \\
\cutinhead{Tidbinbilla, day 148 of 1996}
  256.79 $\pm$   0.02  &  0.79 $\pm$   0.10  &  2.50 &   0.34 &  0.13 &   0.10 $\pm$ 0.03  & -0.6 $\pm$ 0.25\\
  295.00  $\pm$  0.01   & 0.63  $\pm$ 0.02 &   5.68  & 1.18   & 0.21   & 0.34 $\pm$ 0.2 &    No min. \\
\enddata


\tablenotetext{a}{This is a blended feature with two broad components which cannot be resolved.}

\end{deluxetable}

\end{document}